# Room-temperature giant magnetotranstance effect in single-phase multiferroics

Yan-Fen Chang[1,2] and Young Sun[1,2,3]*

[1]Beijing National Laboratory for Condensed Matter Physics, Institute of Physics, Chinese Academy of Sciences, Beijing 100190, China
[2]School of Physical Science, University of Chinese Academy of Sciences, Beijing 100190, China
[3]Center of Quantum Materials and Devices, Chongqing University, Chongqing 401331, China
*youngsun@iphy.ac.cn

Abstract

Single-phase multiferroic materials are usually considered useless because of the weak magnetoelectric effects, low operating temperature, and small electric polarization induced by magnetic orders. As a result, current studies on applications of the magnetoelectric effects are mainly focusing on multiferroic heterostructures and composites. Here we report a room-temperature giant effect in response to external magnetic fields in single-phase multiferroics. A low magnetic field of 1000 Oe applied on the spin-driven multiferroic hexaferrites $BaSrCo_2Fe_{11}AlO_{22}$ and $Ba_{0.9}Sr_{1.1}Co_2Fe_{11}AlO_{22}$ is able to cause a huge change in the linear magnetoelectric coefficient ($\alpha_E = dE/dH$) by several orders, leading to a giant magnetotranstance (GMT) effect at room temperature. The GMT effect is comparable to the well-known giant magnetoresistance (GMR) effect in magnetic multilayers, and thus opens up a door toward practical applications for single-phase multiferroics.

## I. INTRODUCTION

In electric circuits, voltage ($v$), current ($i$), charge ($q$), and magnetic flux ($\varphi$) are four basic circuit variables. The relationships between them have defined three fundamental circuit elements – the resistor with resistance $R=dv/di$, the capacitor with capacitance $C=dq/dv$, and the inductor with inductance $L=d\varphi/di$, as shown in Fig. 1(a). The fourth element defined from the $q$-$\varphi$ relationship was named transtor with transtance $T= dq/d\varphi$ [1], which can be implemented by employing the magnetoelectric (ME) effects, *i.e.*, the mutual control of magnetization ($M$) by electric fields ($E$) and electric polarization ($P$) by magnetic fields ($H$) [2-5]. Each circuit element has a corresponding nonlinear memelement, including the memristor, memcapacitor, meminductor, and memtranstor. Especially, the memristor and memtranstor have been under intensive studies due to their promise in developing next-generation low power electronic devices such as nonvolatile memory, logic gates, and artificial neurons [6-12].

It is well known that the resistance of a resistor may strongly depend on applied dc magnetic fields, yielding various magnetoresistance (MR) effects. For instance, the resistance of some magnetic multilayers and tunneling junctions are very sensitive to external magnetic fields, resulting in the GMR effect [13,14]; a moderate magnetic field of several Tesla can greatly suppress the resistance by several orders in some perovskite manganites, known as the colossal magnetoresistance (CMR) effect [15,16]; some topological materials exhibit non-saturating extreme magnetoresistance (XMR) up to tens of Tesla [17,18]. These diverse MR effects not only provide a playground for fundamental sciences but also serve as a basis for many technological applications. Similarly, the physical quantities of a transtor may also exhibit a notable dependence on external magnetic fields, which would give rise to the magnetotranstance (MT) effects. However, this subject has not been explored yet. In this paper, we report the discovery of a GMT effect at room temperature in single-phase magnetoelectric multiferroics.

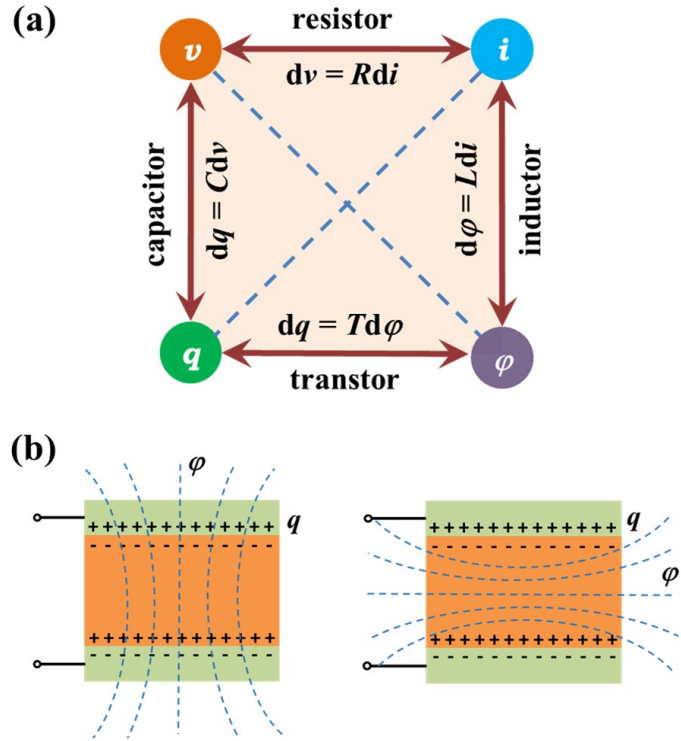

**Figure 1.** (a) The diagram of fundamental circuit elements. The relationships between two of four basic circuit variables (voltage, current, charge, and magnetic flux) define four fundamental elements. (b) Schematic illustration of a transtor with the longitudinal (left) and transverse (right) magnetoelectric effects.

The transtance of a transtor shown in Fig. 1(b) can be obtained by the following equation:

$$T = \frac{dq}{d\varphi} = \frac{CdV}{S'dB} = \frac{S\varepsilon_0\epsilon_r}{S'\mu_0\mu_r} \cdot \frac{dE}{dH} = A\frac{\varepsilon_0\epsilon_r}{\mu_0\mu_r}\alpha_E \quad (1)$$

where $C$ is the capacitance, $V$ is the voltage, $B$ is the magnetic induction, $S$ and $S'$ are the areas of electrodes and magnetic flux, respectively; $\varepsilon_0$ ($\mu_0$) and $\varepsilon_r$ ($\mu_r$) are the permittivity (permeability) of vacuum and the material, respectively; $A=S/S'$ is a geometric factor which is reduced to 1 for the longitudinal configuration; $\alpha_E=dE/dH$ is the linear ME voltage coefficient. Apparently, a $q$-$\varphi$ relationship is set up only if the ME coefficient $\alpha_E$ is non-zero. The calculation of $T$ requires a full assessment of $\varepsilon_r$, $\mu_r$, and $\alpha_E$. To study the influence of external magnetic fields on $T$, one has to measure the magnetic field dependence of $\varepsilon_r$, $\mu_r$, and $\alpha_E$ separately and do the product calculation subsequently, which is quite complicated and inconvenient. Alternately, since the ME coefficient $\alpha_E$ is the characteristic parameter of a transtor, we can focus on its behavior alone, and regard the magnetic field dependence of $\alpha_E$ as a kind of MT effect. In fact, $\alpha_E$ and its magnetic field dependence can be directly measured by a single experiment, which is very important for potential applications. As we show below, the $\alpha_E$ of some single-phase multiferroics strongly depend on the DC bias magnetic field, resulting in a remarkable GMT effect.

## II. MATERIALS AND EXPERIMENTS

The magnetoelectric materials investigated in this work are Y-type hexaferrites which are a prototype of spin-driven multiferroics with spiral magnetic structures [19-24]. In these multiferroic hexaferrites, a small $P$ is usually induced by the transverse conical spin structure in the hexagonal plane, which can be well interpreted by the spin current model [25] or inverse Dzyaloshinskii-Moriya (DM) interaction [26]. Our recent study showed that the Y-type hexaferrites $BaSrCo_2Fe_{11}AlO_{22}$ and $Ba_{0.9}Sr_{1.1}Co_2Fe_{11}AlO_{22}$ exhibit the ME effect even at room temperature [27]. Therefore, we used these samples to test the magnetic field dependence of $\alpha_E$ at room temperature.

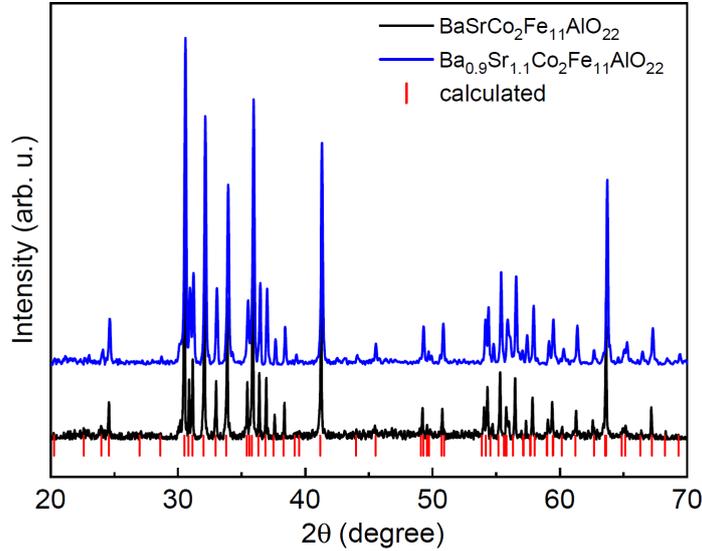

**Figure 2.** Powder x-ray diffraction patterns at room temperature for two Y-type hexaferrite samples.

Polycrystalline samples of $BaSrCo_2Fe_{11}AlO_{22}$ and $Ba_{0.9}Sr_{1.1}Co_2Fe_{11}AlO_{22}$ were synthesized by the solid-state reaction method [24]. Stoichiometric amounts of powders of $BaCO_3$, $SrCO_3$, $Co_3O_4$, $Fe_2O_3$ and $Al_2O_3$ were thoroughly mixed in an agate mortar, and calcinated at 1000 °C for 12 h in air. The calcinated powders were reground and pressed into pellets, then sintered at 1200 °C for 12 h in oxygen atmosphere. Finally, the samples were annealed at 900°C for 72 h in a flow of oxygen in order to reduce the oxygen vacancies and enhance the resistivity. The prepared samples were characterized by powder x-ray diffraction (XRD) at room temperature using a Rigaku x-ray diffractometer. As shown in Fig. 2, all the main diffraction peaks can be indexed to the Y-type hexagonal phase. The XRD results confirm that all the prepared samples are clean single-phase Y-type hexaferrites without detectable impurities.

The magnetic properties were measured by using a Magnetic Properties Measurement System (Quantum Design). The samples were cut into thin plates with the typical size of 3×5×0.2 mm$^3$ for electrical property measurements. Silver paste was painted on two surfaces of the samples to make electrodes. The dielectric permitivity and ME current were measured in a Cryogen-free Superconducting Magnet System (Oxford Instruments, Teslatron PT) using an LCR meter (Aglient 4980A) and an electrometer (Keithley 6517B), respectively. A poling procedure was performed in order to measure the ME current. First, an electric field $E$=1 MV/m and a high magnetic field of 50 kOe (perpendicular to the electric field) were applied together on the samples. Then, the magnetic field was swept down to 5 kOe and $E$ was removed. After the samples were short-circuited for 30 min, the ME currents were recorded while sweeping $H$ down to -50 kOe. $P$ was obtained by integrating the ME current with time.

The ME coefficient $\alpha_E$ was measured by a dynamic method as illustrated in Fig. 4(a). The sample is placed in the middle a solenoid. An AC current source (Keithley 6221) is connected to the solenoid to generate a small AC magnetic field (~ 2 Oe). The induced $V_{ME}$ on the sample is detected by a lock-in amplifier (Stanford Research SR830). The probe is loaded in an Oxford TeslatronPT superconducting magnet system which provides the DC bias magnetic field.

## III. RESULTS AND DISCUSSION

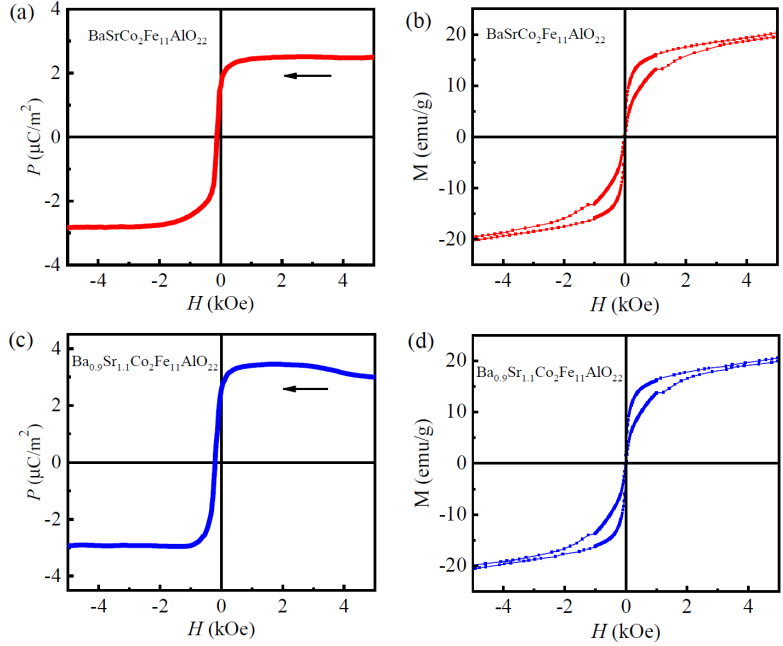

**Figure 3.** Magnetic field reversal of electric polarization at 300 K in (a) BaSrCo$_2$Fe$_{11}$AlO$_{22}$ and (c) Ba$_{0.9}$Sr$_{1.1}$Co$_2$Fe$_{11}$AlO$_{22}$. (b) and (d) The $M$-$H$ loop at 300 K for two samples, respectively. The arrows indicate the direction of sweeping magnetic field.

Figs. 3(a) and 3(b) show magnetic-field control of electric polarization at 300 K for two samples, respectively. One unique feature of these Y-type hexaferrites is that the spin-induced electric polarization can be rapidly reversed by a low magnetic field due to the rotation of conical spin structure with applied magnetic field [19,20]. This leads to the maximum of the ME effect occurring around zero magnetic field. The amplitude of $P$ is about 3 $\mu C/m^2$ at room temperature for both samples, which is three to four orders less than that of traditional ferroelectrics. Therefore, these single-phase multiferroics with room-temperature ME effects are generally considered useless. The insets in Fig. 2 show the *M-H* loops at 300 K. These hexaferrites are soft magnets with a very low coercivity, which is pivotal to the low-field ME effect.

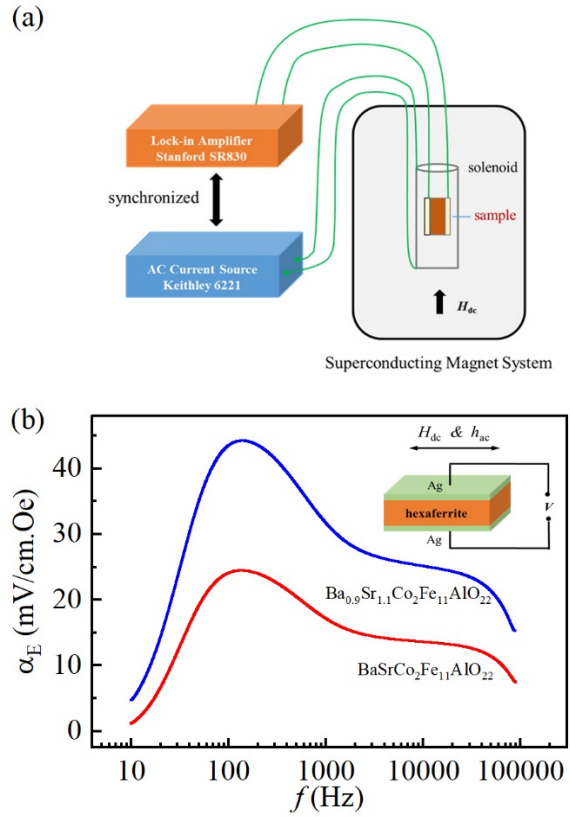

**Figure 4.** Frequency dependence of the ME coefficient $\alpha_E$ at room temperature for two hexaferrite samples. The inset shows the measurement configuration. The direction of DC and AC magnetic fields is perpendicular to the electric field, *i.e.*, in the transverse configuration.

We then focused on the linear ME voltage coefficient $\alpha_E$ of two samples. The measurement configuration is shown in the inset of Fig. 4(b). Both the DC and AC magnetic fields are applied perpendicular to the electric field, *i.e.*, in the transverse configuration, because the spin-induced electric polarization in these hexaferrites is vertical to applied magnetic field. The amplitude of $\alpha_E$ depends on the frequency of AC magnetic field and a broad maximum around 130 Hz is observed, which could be related to the mechanical-resonance enhancement of the ME effect [28].

Fig. 5(a) and 5(b) show $\alpha_E$ as a function of in-plane DC bias magnetic field for two hexaferrite samples, respectively. For the $BaSrCo_2Fe_{11}AlO_{22}$ sample, $\alpha_E$ decreases rapidly from 27 mV/cm.Oe at zero field to 0.24 mV/cm.Oe at 1 kOe and 0.04 mV/cm.Oe at 2 kOe. For the $Ba_{0.9}Sr_{1.1}Co_2Fe_{11}AlO_{22}$ sample, $\alpha_E$ drops from 44 mV/cm.Oe at zero field to 0.52 mV/cm.Oe at 1 kOe and 0.09 mV/cm.Oe at 2 kOe. In other words, a low magnetic field of 2 kOe is able to cause a huge change in $\alpha_E$ by three orders of magnitude.

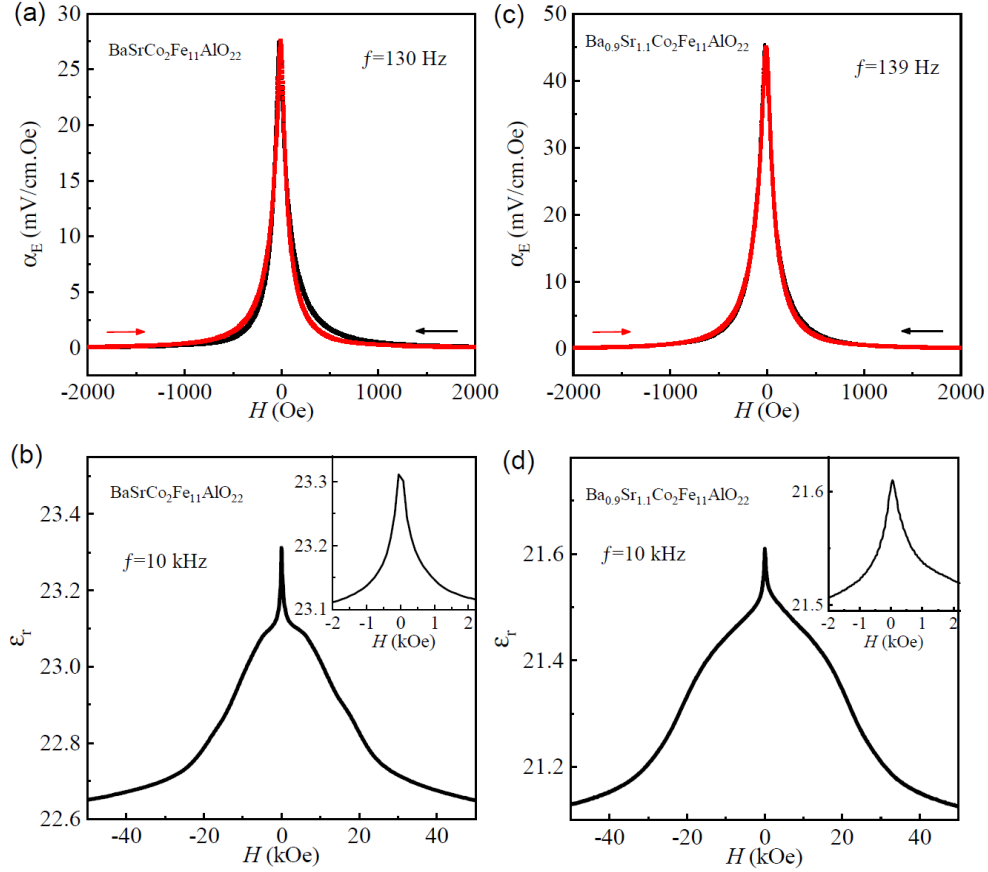

**Figure 5.** The ME coefficient as a function of DC bias magnetic field at 300 K for (a) $BaSrCo_2Fe_{11}AlO_{22}$ and (c) $Ba_{0.9}Sr_{1.1}Co_2Fe_{11}AlO_{22}$. The dielectric permittivity as a function of DC bias magnetic field at 300 K for (b) $BaSrCo_2Fe_{11}AlO_{22}$ and (d) $Ba_{0.9}Sr_{1.1}Co_2Fe_{11}AlO_{22}$. The insets show an expanded view in the low field range. The arrows indicate the direction of sweeping magnetic field.

In comparison, the magnetic field dependence of the dielectric permitivity ($\varepsilon_r$) at 300 K is presented in Figs. 5(b) and 5(c). $\varepsilon_r$ decreases slightly with increasing magnetic field. For $BaSrCo_2Fe_{11}AlO_{22}$, a high magnetic field of 50 kOe causes a relative change ~ 2.8% and a low magnetic field of 2 kOe only induces a change of 0.8% in $\varepsilon_r$. Similarly, for $Ba_{0.9}Sr_{1.1}Co_2Fe_{11}AlO_{22}$, the relative change in $\varepsilon_r$ is ~ 2.2% at 50 kOe and 0.5% at 2 kOe. In fact, the quantity of capacitance (or dielectric permittivity) is generally insensitive to external magnetic field so that a large magnetocapacitance (or magnetodielectric) effect can hardly been observed. In strong contrast, the ME coefficient is much more sensitive to magnetic field than the capacitance.

The MT ratio, defined as MT=[$\alpha_E(H)-\alpha_E(0)$]/$\alpha_E(0) \times 100\%$, is plotted in Figs. 6(a) and 6(b) for two samples, respectively. For BaSrCo$_2$Fe$_{11}$AlO$_{22}$, the MT ratio is as high as 67% at 100 Oe, 96.5% at 500 Oe, and 99% at 1000 Oe. Similarly, for Ba$_{0.9}$Sr$_{1.1}$Co$_2$Fe$_{11}$AlO$_{22}$, the MT ratio reaches 62% at 100 Oe, 95.7% at 500 Oe, and 98.8% at 1000 Oe. We also measured the MT behavior with different frequencies (50 Hz and 1 kHz) of ac magnetic field for BaSrCo$_2$Fe$_{11}$AlO$_{22}$. As seen in Fig. 7(a) and 7(b), although the absolute value of $\alpha_E$ relies on the frequency of AC magnetic field, the GMT effect at room temperature was always observed regardless of the frequency. This giant effect at room temperature is comparable to the well-known GMR effect in magnetic multilayers, and thus can be directly used for practical applications.

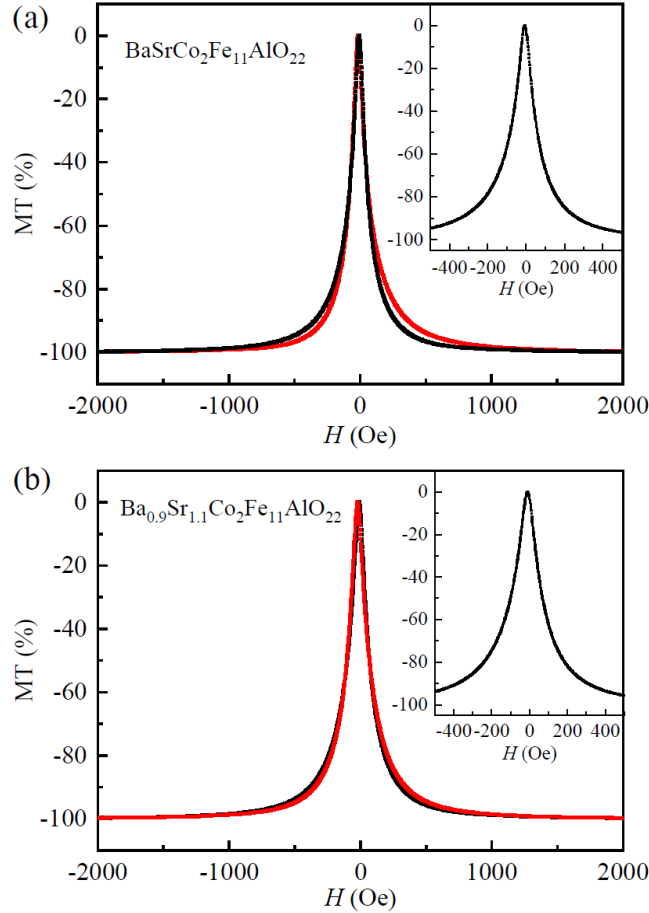

**Figure 6.** The MT ratio defined as MT=[$\alpha_E(H)-\alpha_E(0)$]/$\alpha_E(0)\times100\%$ at room temperature for (a) BaSrCo$_2$Fe$_{11}$AlO$_{22}$ and (b) Ba$_{0.9}$Sr$_{1.1}$Co$_2$Fe$_{11}$AlO$_{22}$. The insets show a zoom-in view in the low field range.

The high sensitivity of the ME coefficient $\alpha_E$ in response to DC bias magnetic field is related to the sharp reversal process of the conical spin structure in multiferroic Y-type hexaferrites [19-24]. As seen in Fig. 3, these hexaferrites are very soft magnets with a low magnetic coercivity. Upon the reversal of magnetization by a small field, the spin-induced electric polarization is also reversed. The ME coefficient is pronounced in the vicinity of zero field because the spin-induced electric polarization changes most fast around zero field (i. e. d$P$/d$H$ is maximum) and then changes slightly with increasing field. It is worthy to note that in some hexaferrites the spin-induced electric polarization does not reverse with magnetic field [24]. Then, the ME coefficient $\alpha_E$ would be minimum rather than maximum around zero field. Although the spin-induced electric polarization is tiny and the absolute value of $\alpha_E$ is not as high as that of magnetoelectric composites [28], the MT ratio is remarkable because $\alpha_E$ is very sensitive to external magnetic field. This phenomenon will totally change the viewpoint on single-phase multiferroics. A large electric polarization and a high ME coefficient are not necessarily required for applications but the sensitivity of the ME coefficient to external fields is more important. In previous studies, a large change of the ME voltage coefficient $\alpha_E$ with DC magnetic field was ever observed in some magnetoelectric composites [8,29]. Nevertheless, little attention has been paid to the magnetotranstance effect in single-phase magnetoelectric multiferroics. Our study represents the first example of GMT effect in the low field regime at room temperature in single-phase multiferroics. Moreover, diverse magnetotranstance effects could exist in many magnetoelectric materials, especially in spin-driven multiferroics.

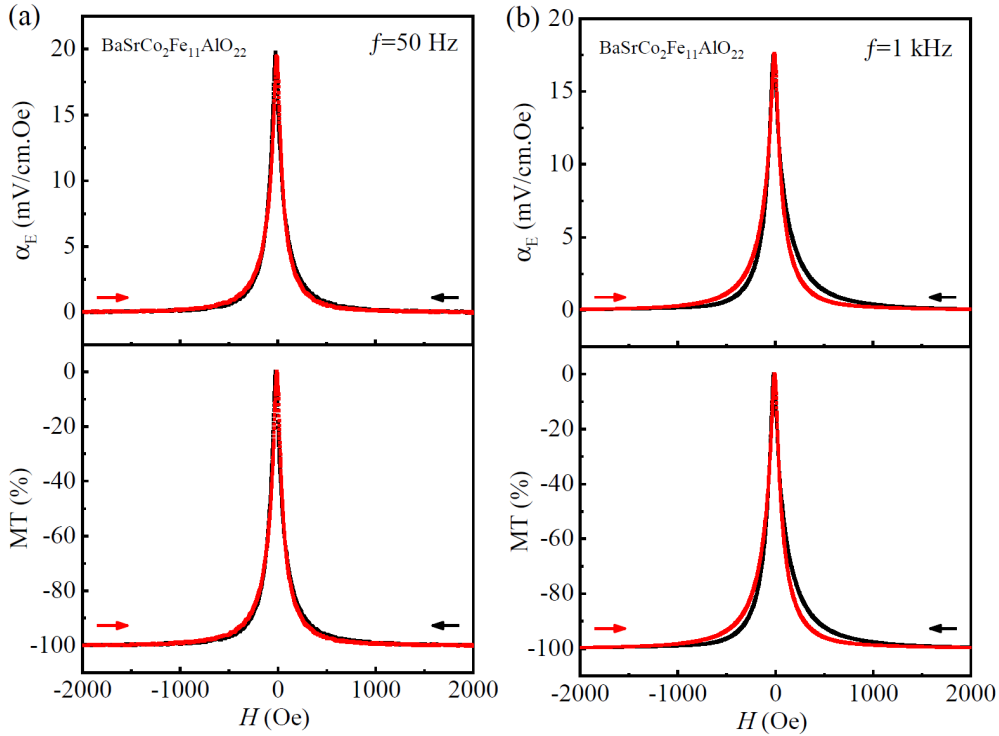

**Figure 7.** The GMT effect at room temperature measured with (a) 50 Hz and (b) 1 kHz AC magnetic field for $BaSrCo_2Fe_{11}AlO_{22}$.

## IV. CONCLUSIONS

In summary, the fourth circuit element "transtor" defined from $T=dq/d\varphi$ is characterized by the ME coefficient $\alpha_E=dE/dH$ which reflects the ability of a material to convert external magnetic field into electric field. It is found that the $\alpha_E$ of some magnetoelectric materials is very sensitive to external DC magnetic field, giving rise to a GMT effect at room temperature. This phenomenon not only holds a promise for technological applications but also opens up a fresh field in condensed matter physics. The variation of $\alpha_E$ of a material or device in response to external stimuli such as temperature, magnetic and electric fields, pressure, light radiation, etc., may give rise to many new phenomena, and would deserve more research efforts in the future.

**ACKNOWLEDGMENTS**

This work was supported by the National Natural Science Foundation of China (Grant Nos. 51725104, 11534015), the National Key Research and Development Program of China (Grant No. 2016YFA0300700), and Beijing Natural Science Foundation (Grant No. Z180009).